\pdfoutput=1

\documentclass[11pt]{article}

\usepackage{ACL2023}

\usepackage{times}
\usepackage{latexsym}

\usepackage[T1]{fontenc}

\usepackage[utf8]{inputenc}

\usepackage{microtype}

\usepackage{inconsolata}
\usepackage{graphicx}
\usepackage{booktabs}
\usepackage{amsmath}

\usepackage{hyperref}

\title{Agentic AI for Human Resources: LLM-Driven Candidate Assessment}

\author{Kamer Ali Yuksel, Abdul Basit Anees, Ashraf Elneima\\\bf{Sanjika Hewavitharana, Mohamed Al-Badrashiny, Hassan Sawaf}\\\\
aiXplain, Inc., San Jose, CA, USA\\
\texttt{\small\{kamer,abdul.anees,ashraf.hatim,sanijka,mohamed,hassan\}@aixplain.com}
}

\begin{document}
\maketitle
\begin{abstract}
In this work, we present a modular and interpretable framework that uses Large Language Models (LLMs) to automate candidate assessment in recruitment. The system integrates diverse sources—including job descriptions, CVs, interview transcripts, and HR feedback—to generate structured evaluation reports that mirror expert judgment. Unlike traditional ATS tools that rely on keyword matching or shallow scoring, our approach employs role-specific, LLM-generated rubrics and a multi-agent architecture to perform fine-grained, criteria-driven evaluations. The framework outputs detailed assessment reports, candidate comparisons, and ranked recommendations that are transparent, auditable, and suitable for real-world hiring workflows. 

Beyond rubric-based analysis, we introduce an LLM-Driven Active Listwise Tournament mechanism for candidate ranking. Instead of noisy pairwise comparisons or inconsistent independent scoring, the LLM ranks small candidate subsets (“mini-tournaments”), and these listwise permutations are aggregated using a Plackett–Luce model. An active-learning loop selects the most informative subsets, producing globally coherent and sample-efficient rankings. This adaptation of listwise LLM preference modeling—previously explored in financial asset ranking \cite{Yuksel2025_LLMListwiseTournaments}—provides a principled and highly interpretable methodology for large-scale candidate ranking in talent acquisition.
\end{abstract}

\blfootnote{HR Manager is online: \href{https://hrmanager.aixplain.com}{https://hrmanager.aixplain.com}}

\section{Introduction}
The process of evaluating job candidates remains one of the most critical and resource-intensive tasks in human resource management. Despite advances in digital recruitment platforms, organizations still struggle to identify and select the most suitable candidates, particularly when faced with high volumes of applicants or highly specialized roles \cite{Raghavan2020,Vaishampayan2025_ResumeMatching}. Current systems tend to over-rely on heuristics, rigid filters, or shallow keyword-based screening, leading to missed opportunities and inconsistent evaluations. 
Human reviewers, while invaluable, often face cognitive overload, inter-rater inconsistencies, and implicit biases that compromise the objectivity and reproducibility of hiring decisions \cite{Quillian2017_MetaHiringDiscrimination}.

To address these limitations, we introduce a robust, LLM-based candidate assessment system designed to enhance the precision, fairness, and scalability of talent evaluations. Our system leverages the generative reasoning capabilities of state-of-the-art language models to generate, apply, and explain fine-grained evaluation criteria specific to each job role. This approach ensures a high degree of alignment between role requirements and candidate evaluations, enabling both depth and breadth in assessments. The system also supports multi-source integration, allowing it to analyze structured and unstructured data from resumes, interviews, recommendation letters, and internal HR notes. Through its modular and extensible architecture, our framework lays the groundwork for a new paradigm in AI-assisted hiring: one that is grounded in interpretability, adaptability, and decision support rather than opaque automation.
\blfootnote{Video demonstration: \href{https://youtu.be/qwcSmWNOHRk}{https://youtu.be/qwcSmWNOHRk}}

Beyond generating structured assessments, our system introduces a new paradigm for ranking candidates: treating the LLM as a listwise fuzzy judge. Rather than comparing two candidates at a time, the model evaluates small groups of candidates simultaneously and produces a relative ordering. These listwise rankings contain richer information per query and mirror how human hiring committees make decisions when reviewing shortlists. We aggregate these orders using a Plackett–Luce model \cite{Plackett1975_AnalysisOfPermutations} to estimate global latent “candidate utilities”, and employ an active-learning \cite{Liang2014_SetwiseActiveLearning} mechanism to adaptively select the most informative candidate groups for evaluation. This approach transforms candidate ranking from an ad-hoc or score-based process into a statistically principled, scalable inference.

\section{Related Work}

Prior efforts in automating candidate assessment have focused primarily on three domains: resume parsing and semantic matching, AI-based video or audio interviews, and the use of predictive analytics to estimate candidate performance. Resume parsing tools such as TextKernel and Sovren have made strides in extracting structured information from unstructured resumes, enabling preliminary filtering. However, these tools often lack contextual understanding and fail to capture the nuances of role-specific competencies \cite{Kanikar2025_ResumeAnalysisReview}. Similarly, semantic matchers like Eightfold and LinkedIn Talent Insights attempt to align job descriptions with candidate profiles using embeddings and cosine similarity measures, but tend to prioritize surface-level alignment over deeper evaluative reasoning \cite{Bevara2025_Resume2Vec}.

Recent developments in LLMs
have enabled few-shot and zero-shot task performance across a wide range of domains. These capabilities have opened up new avenues for generating assessments, feedback, and structured evaluations. However, there remains a significant gap in applying LLMs to the hiring domain in a way that is structured and explainable \cite{ghosh2023jobrecogptexplainablejob}. Our work bridges this gap by building a layered framework that combines LLM-based reasoning with a robust set of evaluation dimensions, customized prompts, and a focus on output format alignment for real-world usability.

Recent work in LLM-based preference learning, including pairwise comparison frameworks such as PAIRS and PoE Bradley–Terry models, demonstrates that LLMs can act as noisy but meaningful preference oracles when aggregated statistically \cite{Qin2024_PRP}. However, nearly all such approaches rely on pairwise comparisons, which are noisy, vulnerable to prompt instability, and do not scale well to large candidate pools \cite{Zeng2024_RankFusion, TrustJudge2025}. Our work is among the first to apply listwise LLM judgments—far more informative than pairwise labels—to candidate assessment. Through listwise tournaments and PL aggregation, we introduce a stable, globally coherent ranking mechanism tailored to HR workflows, bridging the gaps between LLM reasoning, structured evaluation, and hiring decision support.

\section{Architecture}

The proposed system is designed as a multi-stage, modular pipeline where each stage is responsible for a distinct phase in the candidate evaluation lifecycle. At the core of the system is a suite of LLM-powered agents \cite{guo2024largelanguagemodelbased}, each guided by prompt templates and domain-specific heuristics. These agents interact in a sequential yet modular fashion, allowing for plug-and-play flexibility in enterprise settings.

The process begins with the Criteria Generation Agent, which takes as input a job title and description and produces a finely detailed assessment rubric. This rubric is not generic; it is customized to reflect the specific competencies, responsibilities, and context described in the job posting. The rubric includes both technical and non-technical dimensions, and is structured to differentiate candidates not just by qualifications but by demonstrated competencies and growth potential. The system also includes a dedicated Video Question Generation Agent, which tailors reflective and dimension-specific interview questions for video-based responses. Candidate answers are then transcribed and fused with other input modalities to provide a full-spectrum evaluation. 

Next, the Assessment Generator Agent uses this rubric to evaluate candidate profiles. This agent synthesizes data from candidate CVs, HR recommendations, structured interview transcripts, and chat-based conversations. It maps candidate achievements and responses to the rubric and generates a markdown-based report with explicit ratings (Low, Medium, High) for each dimension, supported by detailed justifications. These justifications not only cite specific evidence from inputs but also interpret the relevance of that evidence in the context of the role. To enhance the richness and reliability of assessment, the system integrates multimodal analysis through a video interview analysis module. This component leverages computer vision and audio processing techniques to assess facial expressions, vocal tone, and body language from recorded interviews. These non-verbal cues are fused with the LLM-generated textual insights, offering a holistic picture of a candidate’s interpersonal dynamics, emotional intelligence, and presentation skills—traits often critical in leadership and client-facing roles.

The system also incorporates a Feedback Integration Module that closes the loop between pre-hire assessments and post-hire outcomes. Data such as performance reviews, retention metrics, and team feedback are used to retroactively validate model predictions and refine the scoring logic and rubric generation prompts. This continuous learning mechanism ensures that the system evolves to better predict candidate success, based on empirical evidence. Other agents in the system include a Formatter Agent, which ensures compliance with HR documentation standards; a Comparison Agent, which conducts side-by-side evaluations across top candidates; and a Ranking Agent, which integrates new candidates into existing ranked lists while preserving order consistency. 



\subsection{Prompt Engineering}
Prompt engineering is central to the success of this system \cite{kojima2023largelanguagemodelszeroshot}. Each agent is powered by a dedicated prompt template designed to elicit high-quality, context-sensitive output from the language model. The prompts are carefully structured to ensure that outputs are not only informative but aligned with HR practices.
The Criteria Generation Prompt guides the model to rewrite and enrich default assessment rubrics based on job-specific information. It instructs the model to retain general competencies such as leadership or communication, while augmenting or refining them with role-specific details. The result is a YAML schema that defines evaluation criteria in granular terms, often including 12 to 20 dimensions, each with clear definitions and expectations for various rating scales.

The Assessment Prompt is the heart of the evaluation pipeline. It consolidates inputs from multiple sources—including candidate summaries, HR notes, interviews, and prior assessments—and guides the model to produce a structured report. The prompt enforces a markdown format and mandates explanations for each rating. The prompt also incorporates an internal logic to cross-check dimensions, ensuring coherence and consistency in the assessment. Comparison and Ranking Prompts are designed to simulate the work of hiring committees. They use structured reasoning to evaluate candidates across dimensions and recommend ranked lists, justifying the positioning of each candidate. All prompts are calibrated to avoid hallucination, reduce bias, and maintain a professional tone aligned with corporate communication.

For tournament-based ranking, we design a dedicated Listwise Ranking Prompt that instructs the LLM to evaluate a small group of candidates together, using the job-specific rubric as the evaluation lens. The prompt requires the model to internally reason step-by-step but output only a permutation of candidates from strongest to weakest. This ensures consistency and avoids verbose explanations during the tournament phase. Additional prompts support active learning by incorporating weak prior orderings from the PL model, enabling the LLM to refine its comparisons based on previous rounds while avoiding bias amplification.

\subsection{Evaluation Dimensions}
The assessment criteria generated by the system cover a wide spectrum of attributes required for professional success. These dimensions are not static but tailored dynamically for each job role using the Criteria Generation Agent. Typical categories include domain expertise, technical skill sets, interpersonal effectiveness, leadership capabilities, and career motivation. For example, the Industry-Specific Fit dimension evaluates the extent to which a candidate's experience aligns with the sector targeted by the role. Functional Fit assesses the candidate’s technical proficiency or domain fluency concerning the job’s day-to-day responsibilities. 

Soft skills are also included, with dimensions like Motivation, Drive, and Continuous Learning assessing intrinsic traits and growth mindset. The system ensures that each dimension includes rubric definitions for each level of proficiency. This enables consistent application across candidates and increases inter-rater reliability when used in hybrid human-AI workflows. Listwise tournaments rely on these dimensions as implicit comparison criteria. During tournament ranking, the LLM evaluates subsets of candidates holistically across all rubric dimensions, weighing tradeoffs (e.g., strong technical skill but weaker communication) at a group level. This mirrors human holistic reasoning and produces richer preference signals than independent per-candidate scoring.

\subsection{Report Generation}
The output of the system is designed to be immediately useful to recruiters, hiring managers, and talent committees. Each assessment report is generated in structured markdown format and adheres to a consistent schema that balances detail with readability. Reports begin with a concise introduction outlining the role and the candidate’s context. The core of the report is the dimension-wise assessment, where each dimension is evaluated with a rating and a supporting explanation. These explanations are not generic. They explicitly refer to the candidate’s background, citing projects, metrics, or behavioral indicators found in resumes or interviews. This creates a transparent trail that hiring teams can audit. 

Following the detailed assessment, the system synthesizes an overall readiness evaluation, a cultural fit analysis, and flags areas where the candidate may require support or development. It also suggests alternative roles if the candidate is deemed not optimal for the original role, a feature especially valuable in internal mobility or volume hiring settings. The report concludes with a summary recommendation and a confidence indicator, which can be weighted based on the richness of the input data.

While the listwise tournament mechanism is used to compute global rankings, its outcomes also influence report generation. Candidates’ latent utility values, stability across tournament rounds, and key dimensions driving their placement can be surfaced in the final report, allowing hiring teams to understand why a candidate ranks above peers and to audit the consistency of the evaluation pipeline.

\subsection{Methodology}
To complement rubric-driven candidate evaluation, our system introduces an \emph{active listwise ranking} mechanism that treats the LLM as a fuzzy multi-candidate judge. Instead of independently scoring each candidate or performing unstable pairwise comparisons, the system repeatedly evaluates small groups of candidates using LLM-driven listwise tournaments. This produces richer comparative information and enables a principled, data-efficient approach to building a globally coherent ranking over large applicant pools. The methodology consists of three components: (1) listwise LLM comparisons, (2) probabilistic Plackett--Luce aggregation, and (3) active subset selection via uncertainty.

At each iteration, the system samples a subset of $K$ candidates (typically between 5 and 10) and presents them to the LLM with a structured prompt. The model is instructed to evaluate all candidates \emph{simultaneously}, using the role-specific assessment rubric as the decision lens, perform step-by-step reasoning internally, and output only an ordered list from strongest to weakest fit for the job. The prompt explicitly prohibits justification during these tournament rounds to reduce verbosity and stabilize comparisons. Each listwise ranking encodes $\tfrac{K(K-1)}{2}$ implied pairwise relations and captures cross-dimensional tradeoffs that only emerge when candidates are evaluated side-by-side, imitating human hiring committees.

Each LLM-generated permutation is treated as a noisy observation of latent ``candidate suitability utilities.'' To infer these utilities, we fit a Plackett--Luce (PL) model over all collected tournament results. For each ranking $\pi_t$ from subset $S_t$, the PL likelihood factorizes over positions in the permutation, allowing efficient optimization. The resulting utility vector $\mathbf{u}$ provides a globally consistent ranking across all candidates, stabilizes noise in raw LLM outputs, and reconciles conflicting preferences across tournament rounds. Posterior variances over $u_i$ are approximated via a diagonal Laplace approximation for uncertainty estimates essential in active learning and targeted querying.

Rather than sampling candidate subsets uniformly, we employ an active learning loop that prioritizes the most informative groups. After each PL update, posterior variances highlight candidates with uncertain latent utilities. New subsets are formed to probe contested regions---such as boundaries between ``shortlist'' and ``reject'' tiers. Acquisition strategies, including Monte-Carlo Knowledge Gradient (MC--KG) \cite{balandat2020botorchframeworkefficientmontecarlo}, posterior disagreement sampling \cite{Seung1992_QBC}, and KL-UCB heuristics \cite{Garivier2011_KL_UCB}, determine which subsets maximize expected information gain. This approach sharply reduces the number of LLM queries required and accelerates convergence toward a stable global ranking.

Unlike standalone ranking algorithms, the listwise tournament system utilizes the same role-specific rubric that guides the Criteria Generation and Assessment Generator agents. This ensures that tournament comparisons align with the competency framework used in detailed candidate reports, that latent utilities reflect holistic role fit rather than superficial similarities, and that ranking outcomes remain interpretable and auditable. The final ranking is thus grounded in the same structured criteria used for individual assessments, enabling consistent and transparent decision-making across both comparative and descriptive evaluation modes.

\section{Case Studies}
We applied the system to a set of roles spanning multiple industries and seniority levels to evaluate its generalizability and precision. These included technical roles like AI Research Scientist and Staff Machine Learning Engineer, as well as leadership roles such as VP of Product and CTO for startup environments. Human experts independently rated candidates using the three-level rubric (Low/Medium/High) generated by the Criteria Generation Agent, and the system produced its own ratings using the same rubric. Agreement was computed per dimension, counting cases where the system’s rating was within one level of the human score. Across all evaluations, 87\% of system ratings fell within one band of human scores. In cases where candidates appeared similar, the system was able to surface distinctions that human reviewers later confirmed. 
Although not directly captured by ranking metrics, qualitative inspection of intermediate outputs reveals several desirable behaviors:

\begin{itemize}
    \item Rubric refinement yields clearer, more structured, and more discriminative evaluation.
    \vspace{-0.3cm}\item Subtle distinctions in candidates (e.g., leadership maturity, communication style, depth of reasoning) emerge earlier in iterations.
    \vspace{-0.3cm}\item The system frequently suggests alternative roles for candidates whose strengths are misaligned with the target position. 
\end{itemize}

These findings show that the system not only learns a stable ranking but also produces nuanced, actionable insights for talent evaluation workflows.

\subsection{Experiments}
We evaluate the proposed LLM-driven active listwise tournament framework on a real candidate-ranking task using the full pipeline with iterative criteria refinement. In this configuration, the system not only performs listwise tournaments and Plackett--Luce aggregation but also refines the scoring rubric at each iteration based on LLM critique. The goal of this experiment is to measure ranking fidelity relative to human expert judgments, assess convergence behavior, and characterize how active learning improves ranking stability.

We run 30 iterations of active listwise querying over a fixed pool of candidates. Each consists of:

\begin{enumerate}
    \item Refining the assessment rubric via LLM-based critique.
    \vspace{-0.3cm}\item Selecting next candidate subset using Monte-Carlo Knowledge Gradient (MC--KG).
    \vspace{-0.3cm}\item Obtaining a listwise ranking from the LLM for the selected subset.
    \vspace{-0.3cm}\item Updating global candidate utilities through Plackett--Luce optimization.
\end{enumerate}

\begin{figure}[h]
    \centering
    \includegraphics[width=1.0\linewidth]{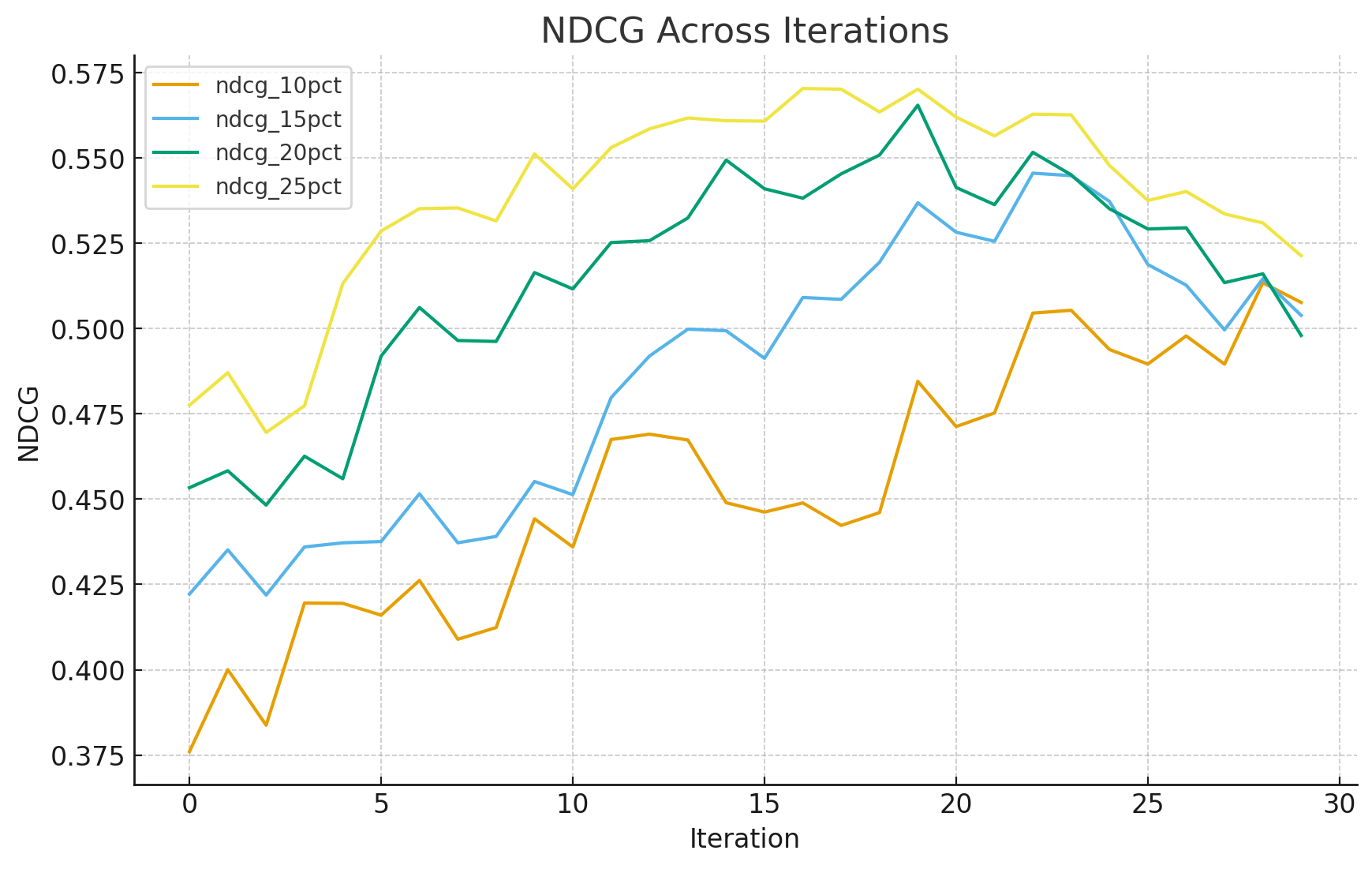}
    \caption{NDCG@K\% progression across 30 iterations.}
    \label{fig:ndcg_over_time}
\end{figure}

We evaluate performance using {NDCG@K} for $K=\{10\%, 15\%, 20\%, 25\%\}$ using human ranking as reference, and convergence metrics that measure the \textbf{stability of the ranking updates}, where higher values mean the ranking is stabilizing and the system is no longer making large structural adjustments. Figure~\ref{fig:ndcg_over_time} shows the evolution of NDCG across cutoffs during the active-learning process.

Across all cutoffs, NDCG improves steadily during the first \textasciitilde20 iterations, with NDCG@25\% achieving the highest value.  
This shows that the active listwise tournament mechanism extracts a ranking structure aligning with human preferences.

\begin{table}[h]
\centering
\caption{Peak NDCG@K values across cut-off ratios.}
\label{tab:peak_ndcg}
\resizebox{\columnwidth}{!}{
\begin{tabular}{lcccc}
\toprule
\textbf{Cutoff} & \textbf{10\%} & \textbf{15\%} & \textbf{20\%} & \textbf{25\%} \\
\midrule
\textbf{NDCG@K} & 0.5134 & 0.5455 & 0.5655 & 0.5703 \\
\bottomrule
\end{tabular}
}
\end{table}

To assess convergence, we track two indicators:

\begin{enumerate}
    \item \textbf{Kendall-$\tau$ between successive iterations}, capturing how similar the ranking at iterations $t$ is to iterations $t-1$.  

    \item \textbf{Utility movement} $\Delta \mathbf{u}$, defined as the norm of the change in the global PL utility vector.
\end{enumerate}

\begin{figure}[h]
    \centering
    \includegraphics[width=1.0\linewidth]{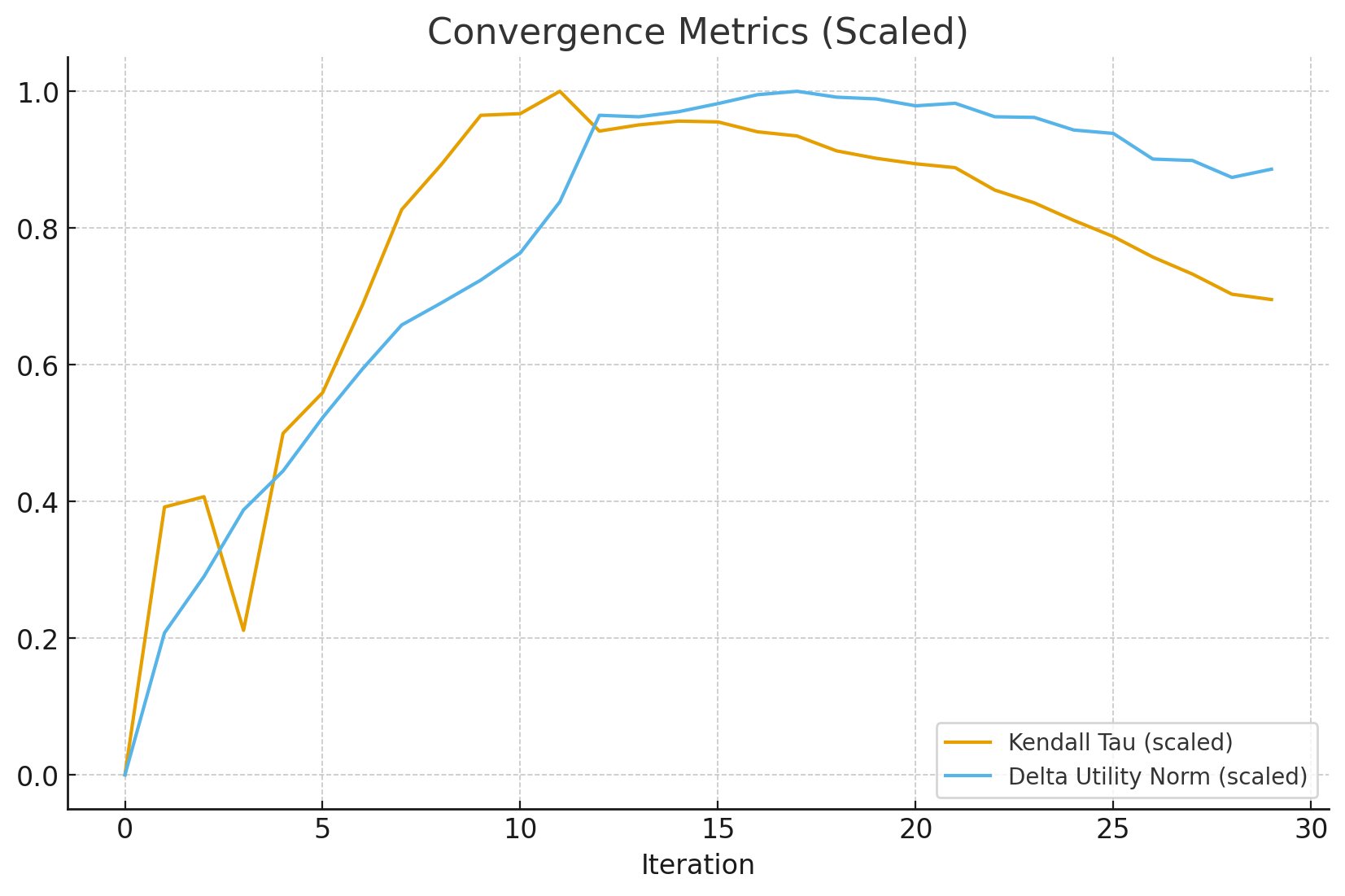}
    \caption{Convergence metrics over 30 iterations. Both Kendall-$\tau$ (stability between iterations) and $\Delta \mathbf{u}$ (norm of utility change) are independently scaled to $[0,1]$.}
    \label{fig:convergence_combined}
\end{figure}

Since these metrics operate on different scales, each is independently normalized to $[0,1]$ for visualization.  
Figure~\ref{fig:convergence_combined} shows their joint progression. The Kendall-$\tau$ curve rises sharply during early iterations, indicating rapid stabilization of the ranking. Meanwhile, $\Delta \mathbf{u}$ decreases substantially after iteration 10, showing that the underlying utility estimates converge. Together, these trends demonstrate that active querying helps the system quickly settle into a stable, high-confidence ranking.

\section{Discussion}
The LLM-based candidate assessment system presents a significant advancement in the intersection of AI and human capital management. By combining rigorous prompt engineering, dynamic rubric generation, and structured reasoning, the system offers a scalable alternative to traditional candidate evaluations. It provides organizations with an interpretable, customizable, and reproducible method for evaluating talent, while also supporting diversity, equity, and inclusion through the criteria.

Nonetheless, there are challenges. The system’s accuracy is influenced by the quality of the input data. Poorly written job descriptions or ambiguous candidate materials can reduce the effectiveness of the evaluation. Additionally, while the system addresses many soft skill dimensions through textual inputs, it can currently recognize facial expressions but cannot yet interpret broader non-verbal cues—such as tone of voice or full-body language. Finally, ongoing prompt tuning is needed to ensure domain-specific relevance in niche or highly specialized roles.

The addition of listwise LLM tournaments introduces a powerful inference layer that helps address a persistent challenge in hiring: ranking candidates with overlapping profiles. By modeling candidate evaluation as a probabilistic ranking problem—rather than independent scoring—we gain stability, interpretability, and resistance to prompt noise. However, this approach also requires careful control of rubric drift and must avoid reinforcing biases across rounds of active querying. These challenges motivate future work combining fairness-aware constraints with PL-model aggregation.

\section{Conclusion}

This paper introduces a novel, LLM-driven candidate assessment system that transforms the way organizations evaluate talent. By uniting the power of language models with structured rubric generation, contextual analysis, and transparent reporting, we offer a framework that is both scalable and interpretable. The system addresses critical limitations in existing hiring workflows by enabling fine-grained, reproducible assessments that align closely with job-specific expectations. With a focus on real-world usability, modular design, and human-centered outputs, our framework lays a strong foundation for the future of AI-assisted hiring.

By extending the Active Listwise Tournament framework from financial asset ranking \cite{Yuksel2025_LLMListwiseTournaments} to candidate evaluation, we show that LLMs can serve not only as rubric-based assessors but also as structured preference oracles. The combination of listwise evaluations, probabilistic PL aggregation, and active learning yields globally coherent candidate rankings that scale to large applicant pools and outperform traditional scoring pipelines. This fusion of LLM reasoning and ranking theory provides a principled foundation for next-generation hiring decision support systems.

Another promising direction is the development of a real-time interview assistant that can suggest probing questions to human interviewers based on live candidate responses. This would create a hybrid workflow where AI augments rather than replaces human judgment, preserving the richness of interpersonal evaluations while enhancing structure and consistency. We also plan to develop a feedback loop that integrates hiring outcomes and performance reviews into the system, enabling it to refine its prompts and scoring models over time.

\newpage

\section*{Limitations}
While the proposed framework offers significant advancements in candidate assessment, it is not without limitations. First, although the integration of LLMs and multimodal analysis allows for nuanced evaluation, the system’s performance is still bounded by the quality and completeness of candidate input data. Sparse or low-quality resumes, brief interview responses, or poorly recorded video inputs can reduce the accuracy of assessments.

Second, although the system supports multilingual evaluation and localization, the depth of cultural adaptation is currently limited to language and structural format. Subtle sociocultural dynamics or communication norms may still be misinterpreted by the LLM or the multimodal modules.

Lastly, while interpretability is a design principle, some of the underlying reasoning from the LLM remains opaque, especially in complex judgment tasks that involve implicit contextual inference. Efforts have been made to structure outputs and provide rationale, but full transparency into model reasoning is still an open challenge.

\section*{Ethics Statement}
This system aims to democratize access to fair candidate assessments, reducing reliance on subjective heuristics and increasing transparency. However, it also raises several ethical considerations.

First, while the system seeks to mitigate bias, LLMs are trained on large-scale internet data that may contain embedded societal biases. If unchecked, these biases can influence rubric generation, assessments, and candidate rankings.

Second, there is the risk of over-reliance on automated systems in critical decision-making. Our system is designed to augment human judgment, not replace it. Recommendations should always be reviewed by qualified HR personnel who contextualize the outputs with domain-specific insight.

Finally, deploying this system at scale across regions and cultures requires sensitivity to local labor laws, cultural practices, and fairness norms.

While these challenges are non-trivial, we believe that the benefits of a well-designed, transparent, and modular AI-assisted evaluation system can meaningfully improve hiring outcomes—if developed and deployed responsibly.

\newpage

\bibliography{anthology,custom}
\bibliographystyle{acl_natbib}


\end{document}